\documentclass[12pt]{article}
\usepackage{amsfonts,amssymb,epsfig,amsmath,mathtools,xcolor}
\usepackage{color}

\usepackage{booktabs}

\usepackage{mathrsfs}
\usepackage{graphicx}
\usepackage{subcaption}
\usepackage{hyperref}
\usepackage{float}
\usepackage{braket}

\renewcommand{\baselinestretch}{1.2}

\newcommand{\eps}{\varepsilon}

\newcommand{\nn}{\nonumber}

\newcommand{\RR}{\mathbb{R}}

\begin{document}

\makeatletter \@addtoreset{equation}{section} \makeatother
\renewcommand{\theequation}{\thesection.\arabic{equation}}
\renewcommand{\thefootnote}{\alph{footnote}}

\begin{titlepage}

\begin{center}
\hfill 
\\

\vspace{2cm}

{\Large\bf Kasner Exponents from the Bouncing Singularity}

\vspace{2cm}

\renewcommand{\thefootnote}{\alph{footnote}}

{\large Hyunwoo Oh}

\vspace{0.7cm}

\textit{Department of Physics and Astronomy \& Center for
Theoretical Physics,\\
Seoul National University, Seoul 08826, Korea.}

\vspace{0.2cm}

E-mail: {\tt hyunwoo1535@snu.ac.kr}

\end{center}

\vspace{0.1cm}

\begin{abstract}

Thermal two-point functions of holographic CFTs can probe the black hole singularity. When a bouncing geodesic exists, the correlator at leading order in $1/N$ develops a singularity $G(t, \vec{x}=0) \propto  (t_c-t)^{-2\Delta+b_t}$, where $\Delta$ is the scaling dimension of the operator. We show that the additional exponent $b_t$ is determined by the Kasner exponents near the singularity. From the wave reflected off the singularity, which gives an exponentially small term in the large-frequency retarded correlator, we derive the formula $b_t=(d-1) \max \left( \frac{p_i-p_t}{1-p_t}, \frac{1}{2} \right)$ for planar black branes in AdS$_{d+1}$, where $p_t$ and $p_i$ are the Kasner exponents of the time and spatial directions. We compare our formula with the numerical values of \v{C}eplak and Valach. It reproduces the relation that they proposed for a special class of black branes. For more general black branes, where no formula had been found, our formula agrees with most of their numerical data and fits them much better than the extension of their relation. We also give predictions for $b_t$ in theories where the Kasner exponents are known, but $b_t$ has not been computed.

\end{abstract}

\end{titlepage}

\renewcommand{\thefootnote}{\arabic{footnote}}

\setcounter{footnote}{0}

\renewcommand{\baselinestretch}{1}

\tableofcontents

\renewcommand{\baselinestretch}{1.2}

\section{Introduction}
\label{sec:intro}

Black hole singularities are hidden behind horizons, but in holography \cite{Maldacena:1997re, Gubser:1998bc, Witten:1998qj}, we can trace the singularity in a boundary quantity \cite{Kraus:2002iv}, the two-point function:
\begin{align}
    G(t,\vec{x}) = \braket{ O(t,\vec{x})\, O(0,\vec{0}) }_\beta ,
\end{align}
where $O$ is a scalar CFT operator of conformal dimension $\Delta$, and $\beta$ is the inverse
temperature. In the large $\Delta$ limit, when analytically continued, $G(t)$ exhibits singularities \cite{Fidkowski:2003nf,Festuccia:2005pi} as
\begin{align} \label{singularityinG}
    G(t) \equiv G(t,\vec{x}=0 ) \propto \frac{1}{ (t-t_c)^{2\Delta } }.
\end{align}
Here $t_c$ is the time of the null geodesic that falls through the horizon, bounces off the black hole singularity and returns to the boundary. For the Schwarzschild black brane in AdS$_{d+1}$, $t_c = \frac{\beta}{2}\cot\frac{\pi}{d} - \frac{i\beta}{2}$ \cite{Fidkowski:2003nf, Amado:2008hw}. The singularity \eqref{singularityinG} is not on the physical sheet: the correlator is analytic in the strip $-\beta < {\rm Im} \, t<0$, which contains $t_c$, and the singularity appears on a secondary sheet of the analytically continued correlator.

At finite $\Delta$, the singularity also can be seen in the OPE. For the Schwarzschild black brane, with the Euclidean time $\tau=it$, the two-point function has the OPE 
\begin{align} \label{eq:expansionofG}
    G(\tau) \equiv G(\tau, \vec{x}=0) = \tau^{-2\Delta} \sum_{n=0}^\infty  \Lambda_n \left(  \frac{\tau}{\beta} \right)^{d n} + \beta^{-2\Delta} \sum_{j=0}^\infty D_j \left( \frac{\tau}{\beta} \right)^{2j} ,
\end{align}
where $d$ is the dimension of the boundary CFT. The coefficients $\Lambda_n$ can be determined from the near-boundary expansion of the bulk field dual to $O$ \cite{Fitzpatrick:2019zqz}, and $D_j$ are constants. We call the first sum $G_T$, the stress-tensor sector of $G(\tau)$, and the second $G_{[OO]}$, the double-trace sector. For other black branes, $G_T$ also contains other operators in the sum, and the powers $dn$ are replaced by their dimensions \eqref{eq:split}.

$G_T(\tau)$ has the singularity at $\tau=\tau_c\equiv i t_c$ \cite{Ceplak:2024bja}, and in the full correlator $G(\tau)$, this is canceled by the double-trace sector $G_{[OO]}(\tau)$ when $\tau$ is in the analytic strip $0<{\rm Re} \,  \tau < \beta$ \cite{Buric:2025anb, Buric:2025fye}. On the secondary sheet, $G_T(\tau)$ and $G_{[OO]}(\tau)$ obtain different phase factors, and the singularity is not canceled so that $G(t)$ reveals the singularity \cite{Komargodski:2026oob}. We parametrize the singularity as
\begin{align} \label{singularityinG_T}
    G_T(t, \vec{x}=0) \sim (t_c-t)^{-(2\Delta -b_t)},
\end{align}
equivalently, $\Lambda_n \sim a^n n^{2\Delta-b_t-1}$ at large $n$. There is an additional power $b_t$ as well as $2\Delta$, which could not be seen in the large $\Delta$ limit \eqref{singularityinG}. The same singularity appears in the retarded correlator $G_{\rm ret}(t, \vec{x}=0 )$. In the frequency space, it corresponds to a term $G_{\rm ret}(\omega)\supset\omega^{q}e^{i\omega t_c}$. The corresponding term at zero momentum, $\omega^{2\Delta-d}e^{i\omega t_c}$, was obtained in the bulk at finite $\Delta$ from the WKB analysis in $d=4$ \cite{Afkhami-Jeddi:2025wra}. It was recovered from the asymptotic quasinormal modes \cite{Dodelson:2025jff}, using the thermal product formula \cite{Dodelson:2023vrw}, and from the monodromy of the bulk wave equation \cite{Jia:2026ryl}. In $d=4$, this term was also checked against the large-$n$ behavior of the multi-stress-tensor OPE coefficients at zero momentum, computed from the near-boundary expansion of the bulk field \cite{Giombi:2026kdz}.

Throughout this paper, $G$ denotes the correlator at leading order in $1/N$, where the singularity is present. At finite Newton constant $G_N$, i.e., finite $N$, the singularity is expected to be resolved due to the quantum corrections. The singularity cannot exist for the Schwarzschild black brane in $d=3,4,5$ \cite{Komargodski:2026oob}, and it is smoothed into a peak of width $\sqrt{G_N}$ in a two-dimensional model \cite{Batra:2026gbq}. Also, \cite{Dodelson:2026hjm}, which appeared on arXiv on the same day as this paper, found evidence that the analogous singularities are smoothed on a time scale $\Delta t\sim N^{-1/2}$ in high-temperature matrix quantum mechanics.

Near the spacelike singularity, the metric of the black brane takes the Kasner form \cite{Kasner:1921zz, Belinsky:1970ew}
\begin{align}
    ds^2=-d\sigma^2+c_t^2\,\sigma^{2p_t}\,dt^2+c_1^2\,\sigma^{2p_i}\,d\vec x^{\,2},\qquad \sigma\to0,
\end{align}
where $\sigma$ is the proper time to the singularity, $t$ and $\vec x$ are the boundary coordinates. $c_t,c_1$ are constants determined from the original black brane metric. \cite{Frenkel:2020ysx} discovered that the Kasner exponents were interesting quantities to probe in holography. In the large-$\Delta$ limit, they showed that the Kasner exponent $p_t$ determines a non-analytic correction $\propto |t-t_c|^{-1/p_t}$ to the geodesic length near the bouncing singularity. The Kasner exponents are also encoded in the high-overtone quasinormal modes. They govern the momentum (or angular-momentum) dependence of the subleading corrections at large overtone number \cite{Hartnoll:2026vhu, Coviello:2026tol, Coviello:2026hdc, Shi:2026cfq}. Combined with the probe-mass dependence, which involves only $p_t$, this determines $p_t$ and $p_i$ independently \cite{Xiao:2026pir}. Recently, \cite{Ceplak:2026ard} observed $b_t$ numerically using the large-$n$ behavior of $\Lambda_n$, and proposed that $b_t$ may be determined from the Kasner exponents, for a special case.

In this paper, we derive the formula of $b_t$ as
\begin{align} \label{b_t:general}
    b_t  = (d-1) \, \max \left( \frac{p_i -p_t}{1-p_t}, \frac12  \right) ,
\end{align}
in terms of the Kasner exponents. We consider spacelike singularities, set all $d-1$ spatial directions to have the same exponent $p_i$, and assume $p_i, p_t < 1$. We do not assume the vacuum conditions $p_t+(d-1)p_i=p_t^2+(d-1)p_i^2=1$, which hold for pure gravity. With the exponent $\alpha \equiv  \frac{2(1-p_i)}{1-p_t} $ in \cite{Hartnoll:2026vhu} (denoted $q_m$ in \cite{Coviello:2026tol, Coviello:2026hdc}), we have $\frac{p_i-p_t}{1-p_t}=1-\frac{\alpha}{2}$ and \eqref{b_t:general} can be expressed as $b_t=(d-1) \max \left(1-\frac{\alpha}{2}, \frac{1}{2} \right)$. Therefore, $b_t$ is determined by the Kasner region when $\alpha<1$, and determined only by the spacetime dimension when $\alpha \geq 1$. Our formula reproduces $b_t=\frac d2$ for the Schwarzschild black brane \cite{Jia:2026pmv}.

For the case where the $d-1$ spatial directions have different exponents $p_i$, we conjecture the more general formula:
\begin{align}
    b_t=\sum_{j=1}^{d-1}\max\left(\frac{p_j-p_t}{1-p_t},\frac12\right).
\end{align}

The paper is organized as follows. In Section~\ref{sec:2}, we relate the singularity of $G_T(\tau)$ to that of the retarded correlator $G_{\rm ret}(t)$. We explain that the singularity of $G_{\rm ret}(t)$ comes from the exponentially small term in $G_{\rm ret}(\omega) $ at large frequency $\omega$. In Section~\ref{sec:3}, following the proposal of \cite{Afkhami-Jeddi:2025wra}, we explain how the exponentially small term of the retarded correlator at large $\omega$ arises from a wave reflected at the singularity, and how we choose this wave. In Section~\ref{sec:derivation}, we derive the formula \eqref{b_t:general}. We keep the spatial momentum $\vec k$ and integrate over $\vec k$ to reach $G_{\rm ret}(t,\vec x=0)$. In Section~\ref{sec:5}, we compare our formula with the numerical results of \cite{Ceplak:2026ard}. For the black branes for which they proposed their relation, $p_i-p_t=1$ holds and \eqref{b_t:general} reduces to their proposal $b_t=\frac{d-1}{1-p_t}$. In general, however, $b_t$ depends on $p_i$ as well as on $p_t$. For black branes with $p_i-p_t\neq1$, for which no formula had been found, our formula agrees with most of their data, and it fits them much better than the extension $b_t=\frac{d-1}{1-p_t}$ of their relation. In Section~\ref{sec:6}, we predict $b_t$ from the Kasner exponents, in the Einstein--scalar and Einstein--Maxwell--dilaton black branes in AdS$_4$. In these theories, $b_t$ has not been numerically computed. In Section~\ref{sec:7}, we conclude with remarks and future directions.

\section{From the OPE to the large-frequency retarded correlator}
\label{sec:2}

In a holographic CFT at large $N$ and strong 't Hooft coupling, the thermal two-point function $G(\tau)=\braket{O(\tau)O(0)}_\beta$ at $\vec x=0$ splits into two sectors \cite{Ceplak:2024bja},
\begin{align} \label{eq:split}
    G(\tau) &= G_T(\tau)+G_{[OO]}(\tau), \nn\\
    G_T(\tau) &= \frac{1}{\tau^{2\Delta}}\sum_{n=0}^\infty\Lambda_n\left(\frac{\tau}{\beta}\right)^{s_n}, \qquad
    G_{[OO]}(\tau) = \frac{1}{\beta^{2\Delta}}\sum_{j=0}^\infty D_j\left(\frac{\tau}{\beta}\right)^{2j}.
\end{align}
This follows from the OPE $O(\tau) O(0) = \sum_{O'} C_{O'} \, \tau^{\Delta'-2\Delta} \, O'(0)$, schematically with spin indices suppressed, where $\Delta'$ is the dimension of $O'$. Since $\braket{O'}_\beta\propto\beta^{-\Delta'}$, each operator contributes a term $\tau^{-2\Delta}(\tau/\beta)^{\Delta'}$. The double-trace operators have dimensions $\Delta'=2\Delta+2n+\ell$, so their contributions are $\beta^{-2\Delta}(\tau/\beta)^{2n+\ell}$, and only those with even spin $\ell$ have non-zero thermal expectation values. This gives $G_{[OO]}$. The remaining operators, whose dimensions $s_n$ do not depend on $\Delta$, give $G_T$. They are the identity, the stress tensor and its composites, and, in backgrounds with matter fields, also the operators dual to these fields and their composites.

For the Schwarzschild black brane, $s_n=dn$ \cite{Ceplak:2024bja}. For the black branes \eqref{eq:metric} in even $d$, \cite{Ceplak:2026ard} takes $G_T$ to be the ``current sector'' and $s_n=2n$. When a bouncing geodesic exists, $G_T$ has the bouncing singularity as
\begin{align} \label{eq:defb}
    G_T(\tau ) \sim (\tau_c-\tau)^{-(2\Delta-b_t)},
\end{align}
with $\tau_c = it_c$. The retarded correlator is given by
\begin{align}\label{eq:Gretdef}
    G_{\rm ret}(t, \vec x)= i \Theta(t) \braket{ [O(t,\vec x), \, O(0,\vec 0)] }_\beta.
\end{align}
We let $G_{\rm ret}(t) \equiv G_{\rm ret}(t, \vec{x}=0)$. This is the difference of the two Wightman functions $\braket{O(t)O(0)}_\beta = G(it+0^+)$ and $\braket{O(0)O(t)}_\beta = \braket{O(-t)O(0)}_\beta = G (-it+0^+)$,
\begin{align} \label{eq:totaldiffer}
    G_{\rm ret} (t) = i\Theta (t) \left[ G(it+0^+) - G(-it+0^+)  \right] .
\end{align}
At $\tau=\pm it+0^+$, we evaluate the powers of $\tau$ in \eqref{eq:split} on the principal branch, $(\pm it)^{a}=t^{a}e^{\pm i\pi a/2}$, which is their continuation from real $\tau>0$ within ${\rm Re}\,\tau>0$. For small $t$, both sums in \eqref{eq:split} converge, so they can be evaluated term by term. In \eqref{eq:totaldiffer}, the double-trace sector drops out, since $\tau^{2j}$ takes the same value at $\tau=\pm it$, while $\tau^{-2\Delta+s_n}$ acquires the different phases $e^{\pm i\pi(s_n/2-\Delta)}$. Hence, we obtain
\begin{align}\label{eq:GretGT}
    G_{\rm ret}(t) &= i\Theta(t) \left[ G_T(it+0^+) - G_T(-it+0^+)\right] \nn\\
    &= 2\,\Theta(t)\, t^{-2\Delta}\sum_{n=0}^\infty \sin\left[\pi\left(\Delta-\frac{s_n}{2}\right)\right]\Lambda_n\left(\frac{t}{\beta}\right)^{s_n} ,
\end{align}
for small $t>0$. For larger and complex $t$, we drop $\Theta(t)$ and define $G_{\rm ret}(t)$ by the analytic continuation of the right-hand side. When all $s_n$ are even, as for the Schwarzschild black brane in even $d$, we obtain $\sin[\pi(\Delta-\frac{s_n}{2})]\,t^{s_n}=\sin(\pi\Delta)\,(it)^{s_n}$ and \eqref{eq:GretGT} becomes
\begin{align}\label{eq:GretGTeven}
    G_{\rm ret}(t) =  2\sin(\pi\Delta)\,\Theta(t)\, t^{-2\Delta }\sum_{n=0}^\infty\Lambda_n \left(\frac{it}{\beta}\right)^{s_n} = 2 \sin (\pi \Delta)\, e^{i\pi \Delta }\,\Theta(t)\, G_T( \tau=it).
\end{align}
So, for generic $\Delta$, $G_{\rm ret}(t)$ has the same singularity at $t=t_c$ as $G_T(\tau=it)$, with the same power,
\begin{align}\label{eq:Gretsing}
    G_{\rm ret}(t)\sim(t_c-t)^{-(2\Delta-b_t)}. 
\end{align}
For other $s_n$, the term $G_T(-it+0^+)$ in the first line of \eqref{eq:GretGT} is evaluated near $\tau=-it_c$. We assume that it neither produces a stronger singularity at $t=t_c$ nor cancels the singularity of $G_T(it+0^+)$, so that \eqref{eq:Gretsing} still holds.

Since $G_{\rm ret}(t)$ is real for real $t$, it is also singular at the complex conjugate point $t_c^*$  with the same exponent. This is the point in the upper half-plane ${\rm Im} \,t>0 $, which produces the exponentially small term of $G_{\rm ret}(\omega)$ at large positive $\omega$, as shown below. From now on, we denote it by $t_c$, so that ${\rm Im}\,t_c = \frac{\beta}{2} >0$.\footnote{The limit $\omega\to-\infty$ picks instead the original point, now $t_c^*$ with ${\rm Im} \,  t_c^* <0$ in the lower half-plane, and gives the same exponent, since $G_{\rm ret}(-\omega)=G_{\rm ret}(\omega)^*$. Working with $\omega\to+\infty$ is only a convention.} For example, we set $t_c= \frac{\beta}{2} \left( \cot \frac{\pi}{d} +i \right)$ for the Schwarzschild black brane.

The singularity of $G_{\rm ret}(t)$ at $t_c$ is encoded in the large-$\omega$ behavior of $G_{\rm ret}(\omega,\vec x=0)=\int_0^\infty dt\,e^{i\omega t}\,G_{\rm ret}(t)$. Each power of $\omega$ in its large-$\omega$ expansion gives a singularity only at $t=0$. As in \cite{Afkhami-Jeddi:2025wra}, a singularity at $t_c$ comes instead from a term
\begin{align}\label{eq:q}
    G_{\rm ret}(\omega,\vec x=0)\supset C\,\omega^{q}\,e^{i\omega t_c},\qquad \omega\to+\infty,
\end{align}
since $\int^\infty d\omega\,\omega^{q}e^{-i\omega(t-t_c)}\propto(t_c-t)^{-q-1}$. Comparing with \eqref{eq:Gretsing}, one can find $b_t=2\Delta-q-1$. Since ${\rm Im}\,t_c=\beta/2$, this term is exponentially small, $|e^{i\omega t_c}|=e^{-\beta\omega/2}$. This does not appear in the expansion in powers of $1/\omega$. To determine $b_t$, it is enough to evaluate this term.

\section{Steepest-descent solution and the reflected wave}
\label{sec:3}

In this section, we set up the wave equation in the black brane and explain how the exponentially small term of $G_{\rm ret}(\omega, \vec{k})$ arises.

We consider a planar black brane in AdS$_{d+1}$,
\begin{align} \label{eq:brane}
    ds^2=-r^2f(r)\,dt^2+\frac{dr^2}{r^2h(r)}+r^2d\vec x^{\,2}_{d-1},
\end{align}
where $f,h\to1$ at the boundary $r\to\infty$. We assume that $f$ and $h$ have a common simple zero at the horizon $r=r_h$, are positive for $r>r_h$, and are negative for $0<r<r_h$, so that the singularity at $r=0$ is spacelike. The inverse temperature is $\beta= 4\pi / (r_h^2 \sqrt{f'(r_h) h'(r_h)} )$.

We use the tortoise coordinate $r_*$, which is defined by $\frac{dr^2}{r^2 h(r)} = r^2 f(r) dr_*^2$,
\begin{align} \label{eq:rstar}
    r_*(r)=\int_0^r\frac{dr'}{r'^2\sqrt{fh}} ,
\end{align}
where the integral constant was fixed so that $r_*(0)=0$. We set $\sqrt{fh}>0$ outside the horizon, continued analytically through its zero at $r_h$. Inside the horizon $r<r_h$, we set $\sqrt{fh}<0$, so $r_*<0$ there. The integrand has a pole at $r_h$, which we pass below the real axis, as in \cite{Afkhami-Jeddi:2025wra}.

The coordinate time for a radial null ray, which travels from the boundary to the singularity and back, is twice the value of $r_*$ at the boundary,
\begin{align} \label{eq:tc}
    t_c= 2 \, r_*(\infty) = 2\int_0^\infty \frac{dr}{r^2\sqrt{fh}} .
\end{align}
In Section~\ref{sec:phase}, we will see that the wave reflected at the singularity indeed carries the factor $e^{i\omega t_c}$, so that $G_{\rm ret}(t)$ is singular at $t=t_c$. The integrand is real on the real $r$ axis except at the horizon $r=r_h$, where
\begin{align}
    r^2\sqrt{fh} \, & \approx \,  r_h^2  \sqrt{f'(r_h)h'(r_h)} \, (r-r_h) \,  = \,  (4\pi/\beta)(r-r_h) .
\end{align}
The contour passes below this pole, and the half--residue gives ${\rm Im} \, t_c = \beta/2$.

A bulk scalar $\Phi=e^{-i\omega t+i\vec k\cdot\vec x}\phi(r)$ of mass $m$, with $m^2 =\Delta (\Delta -d)$, is dual to the operator $O$ in the boundary CFT. In the background \eqref{eq:brane}, the Klein--Gordon equation $\nabla^2\Phi=m^2\Phi$ becomes
\begin{align}
    \frac{1}{r^{d-1}\sqrt{f/h}}\frac{d}{dr}\left(r^{d+1}\sqrt{fh}\,\frac{d\phi}{dr}\right)+\left(\frac{\omega^2}{r^2f}-\frac{k^2}{r^2}-m^2\right)\phi=0 .
\end{align}
With $\phi(r) \equiv r^{-(d-1)/2}\Psi(r)$ and the tortoise coordinate $r_*$, this wave equation can be written in the Schr\"odinger form as
\begin{align} \label{eq:schr}
    \frac{d^2 \Psi }{ dr_*^2 } +  \left( \omega^2-k^2f-V_0 \right) \Psi=0,
\end{align}
where
\begin{align} \label{eq:V0}
    V_0=m^2r^2f+\frac{d-1}{4}\left[(d+1)\,r^2fh+r^3\,(fh)'\right] 
\end{align}
does not depend on $k$ or $\omega$.

By the holographic dictionary, the retarded Green's function $G_{\rm ret}(\omega,\vec k)$ is, up to a $\Delta$-dependent normalization, given by the ratio of the coefficients of $r^{-\Delta}$ to that of $r^{\Delta-d}$ in the near-boundary expansion of the infalling solution \cite{Son:2002sd}. As we explained in Section~\ref{sec:2}, we need to compute its exponentially small term at large $\omega$. To compute it, following the proposal of \cite{Afkhami-Jeddi:2025wra}, we use the steepest-descent solution $\Psi_{\rm sd}$ defined below, instead of the infalling solution. The proposal which we adopt is that the Green's function computed from $\Psi_{\rm sd}$ is the canonical Borel sum of the large-$\omega$ expansion of $G_{\rm ret}$.

\begin{figure}[H]
    \centering
    \includegraphics[width=\textwidth]{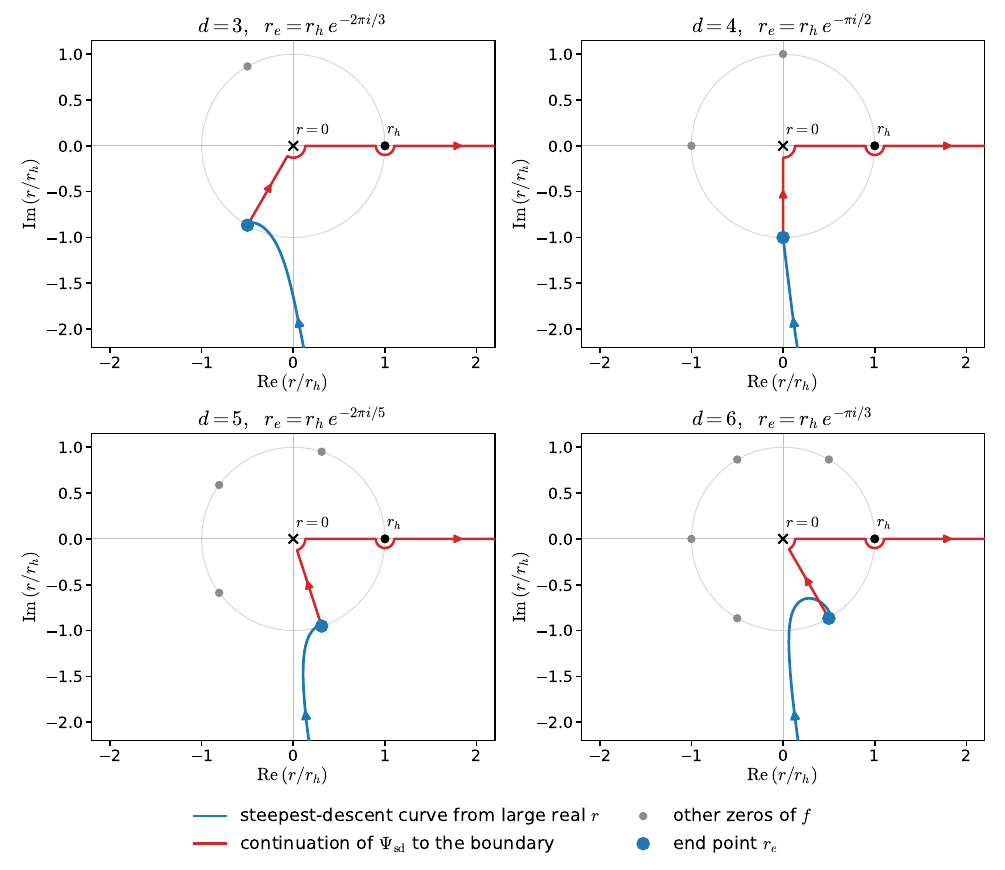}
    \caption{Steepest-descent curves of the infalling wave $e^{-i\omega r_*}$ in the complex $r$ plane for the Schwarzschild black brane in AdS$_{d+1}$, $d=3,\dots,6$. They are shown in blue curves, started at $r=30 \, r_h$, and end at the complex zero $r_e$ of $f$. Red curve shows the path along which $\Psi_{\rm sd}$ is continued to the boundary, along a ray from $r_e$ to the singularity $r=0$, then along the real axis, passing below the horizon $r=r_h$.}
    \label{fig:steepest}
\end{figure}

At large $\omega$, the infalling solution of \eqref{eq:schr} becomes $e^{-i \omega r_*}$. Starting from large real $r$, we continue the solution into the complex $r$ plane along the curve where $e^{-i\omega r_*}$ decreases fastest, which means ${\rm Re} \, r_*$ is constant and ${\rm Im} \, r_*$ decreases. If $f=h$, $r_*$ diverges logarithmically at a complex zero of $f$, and the curve can end there. For the Schwarzschild black brane, $f=h=1-(r_h/r)^d$, we trace the steepest-descent curve numerically (Fig.~\ref{fig:steepest}). It ends at $r_e=r_he^{-2\pi i/d}$ for $3\le d\le6$.

Near the end point $r=r_e$, we have $r_*\simeq c\log(r-r_e)$ with $c \equiv 1/\left(r_e^2f'(r_e)\right)$, so the two solutions behave as $e^{\mp i\omega r_*}\simeq(r-r_e)^{\mp i\omega c}$. Since ${\rm Im}\,r_*$ decreases along the curve and the curve ends at $r=r_e$, we have ${\rm Im}\,r_*\to-\infty$ there. Writing $r-r_e=\rho\,e^{i\vartheta}$ and using that ${\rm Re}\,r_*$ is constant along the curve, we find ${\rm Im}\,r_*\simeq\frac{|c|^2}{{\rm Im}\,c}\log\rho$, so ${\rm Im}\,r_*\to-\infty$ as $\rho\to0$ requires ${\rm Im}\,c>0$. Since $|e^{\mp i\omega r_*}|=e^{\pm\omega\,{\rm Im}\,r_*}$, the continuation of the infalling wave, $e^{-i\omega r_*}$, vanishes at $r_e$, while $e^{+i\omega r_*}$ diverges there. We define $\Psi_{\rm sd}$ as the solution that vanishes at $r_e$, and this is unique up to normalization. This plays the role of the infalling condition at the real horizon $r_h$, which is also a zero of $f$, with $r_*\simeq c_h\log(r-r_h)$. There $c_h \equiv 1/(r_h^2 f'(r_h))=\beta/4\pi$ is real, so neither solution $(r-r_h)^{\mp i\omega c_h}$ vanishes at $r_h$, and we must instead pick the infalling one, $e^{-i\omega r_*}$, as required for the retarded correlator.

However, for black branes with $f\neq h$, the complex zeros of $f$ and $h$ do not coincide in general. Near a simple zero $r_0$ of $f$ at which $h\neq0$, the integrand of \eqref{eq:rstar} behaves as $1/(r^2\sqrt{fh})\propto(r-r_0)^{-1/2}$, which is integrable. Thus, $r_*$ stays finite there and $r_0$ is a square-root branch point rather than an end point. We therefore need a more general prescription.

We take it from the Schwarzschild black brane, where the curve has an end point. As we show in Section~\ref{sec:reflection} for $3\le d\le6$, the solution $\Psi_{\rm sd}$ that vanishes at $r_e$, decays near the singularity as $|y|\to\infty$ along rays from $y=0$ in the sector $-2\pi<\arg y<-\pi$, where $y=\omega x$ is the variable of Section~\ref{sec:reflection}, positive on the real $r$ axis inside the horizon. This condition refers only to the region near the singularity, and it fixes the solution uniquely up to normalization. So we take this as our prescription, and use it to define $\Psi_{\rm sd}$ in general.

Near the singularity, the wave equation has two solutions, $\Psi_{\rm in}$ and $\Psi_{\rm out}$, whose WKB forms are $e^{-i\omega r_*}$ and $e^{+i\omega r_*}$, respectively. Thus $\Psi_{\rm in}$ continues the infalling wave into the interior, while $\Psi_{\rm out}$ continues to the outgoing wave at the boundary. The solution $\Psi_{\rm sd}$ selected in this way is a combination
\begin{align}\label{eq:R}
    \Psi_{\rm sd} \propto \Psi_{\rm in} - R\, \Psi_{\rm out} .
\end{align}
We call $R$ the reflection coefficient at the singularity: the wave sent towards the singularity comes back with amplitude $R$. This is the counterpart of the null bouncing geodesic in the wave picture. In Section~\ref{sec:reflection}, following \cite{Hartnoll:2026vhu}, we show that $R$ depends on the momentum $\vec k$ only through the combination $k^2/\omega^\alpha$, where $\alpha \equiv \frac{2(1-p_i)}{1-p_t}$. We compute $R$ at $\vec{k}=0$, but we do not obtain $R$ explicitly for $\vec k\neq0$.

We then continue $\Psi_{\rm sd}$ along the real $r$ axis to the boundary. On the way, $\Psi_{\rm in}$ and $\Psi_{\rm out}$ do not mix, but they acquire different phases. We denote by $S(\omega,k)$ the phase of $\Psi_{\rm out}$ relative to $\Psi_{\rm in}$ at the boundary. Near the boundary $r\gg r_h$, we have $f\simeq h\simeq1$, and the wave equation reduces to Bessel's equation. In terms of the Hankel functions, $\phi_{\rm sd}=r^{-(d-1)/2}\Psi_{\rm sd}$ is
\begin{align} \label{eq:sdbdy}
    \phi_{\rm sd} \propto r^{-d/2} \left[ H^{(1)}_{\Delta-\frac d2} \left( \frac{\sqrt{\omega^2-k^2}}{r} \right)-R\,e^{iS}\,H^{(2)}_{\Delta-\frac d2} \left(\frac{\sqrt{\omega^2-k^2}}{r} \right) \right],
\end{align}
where $H^{(1)}$ and $H^{(2)}$ continue $\Psi_{\rm in}$ and $\Psi_{\rm out}$, and the constant phases from their normalization are absorbed into $S$. In Section~\ref{sec:phase}, we compute this total phase $S$ and find $S = \omega t_c + \theta \frac{k^2}{\omega} +  \cdots$. Since ${\rm Im}\,t_c=\beta/2$, the factor $e^{iS}$ is exponentially small at large $\omega$.

As one can see in \eqref{eq:sdbdy}, the infalling solution $H^{(1)}$ differs from $\Psi_{\rm sd}$ only by the exponentially small second term, so by the proposal of \cite{Afkhami-Jeddi:2025wra}, the exponentially small term of $G_{\rm ret}$ comes from this term. This gives the term of $G_{\rm ret}(\omega,\vec k)$,
\begin{align} \label{eq:Gbounce}
    G_{\rm ret}(\omega,\vec k)\supset\mathcal N\,(\omega^2-k^2)^{\Delta-\frac d2}\,e^{iS(\omega,k)}\,R ,
\end{align}
where $\mathcal N$ is a non-zero constant independent of $\omega$ and $k$. In Section~\ref{sec:integral}, we perform the momentum integral to obtain $G_{\rm ret}(\omega, \vec{x}=0)$ and $b_t$.

\section{Derivation of $b_t$}
\label{sec:derivation}

We now compute $R$ and $S$ in \eqref{eq:Gbounce}, in Sections~\ref{sec:reflection} and \ref{sec:phase}. Then, we do the momentum integral to obtain $b_t$ in Section~\ref{sec:integral}.

\subsection{Reflection coefficient near the singularity}
\label{sec:reflection}

Near the singularity ($r \ll r_h$), the black brane metric \eqref{eq:brane} reduces to the following Kasner metric:
\begin{equation}
    ds^2 = - d\sigma  ^2+c_t^2 \sigma ^{2p_t} dt^2+c_1^2 \sigma^{2p_i} d \vec x^2 , \qquad   \sigma \to 0,
\end{equation}
with one isotropic exponent $p_i$ for the $d-1$ spatial directions. We assume $p_i,p_t<1$. Comparing with \eqref{eq:brane}, $r=c_1\sigma^{p_i}$, $-r^2f=c_t^2\sigma^{2p_t}$ and $d\sigma = dr/( r\sqrt{-h} )$. Since $\sqrt{fh}<0$ inside the horizon, $dr_* = -d \sigma / (c_t \sigma^{p_t})$, so the coordinate $x \equiv -r_*$ is
\begin{align}
    x=\int_0^\sigma\frac{d\sigma'}{c_t\sigma'^{p_t}}=\frac{\sigma^{1-p_t}}{(1-p_t)c_t},
\end{align}
from \eqref{eq:rstar}. $x$ is finite and goes to $0$ at the singularity, since $p_t<1$. The Kasner region condition $r \ll r_h$ becomes $x\ll \frac{1}{r_h}$ for $x$. The wave equation in the Kasner region is \eqref{eq:schr} with $r_*=-x$,
\begin{equation} \label{waveeqforx}
    \Psi'' + \left(\omega^2 -  V \right) \Psi=0 , \qquad V = V_0+k^2f,
\end{equation}
where $'=d/dx$. Evaluating $V_0$ in \eqref{eq:V0} and $f$ with the relations above and $P\equiv(d-1)p_i$, we obtain
\begin{align}
    V =  -  \frac{ P (2-2p_t -P) }{4 x^2(1-p_t)^2 }   - \frac{c_t^2}{c_1^2} k^2  \left[ (1-p_t ) c_t x   \right]^{2(p_t-p_i)/(1-p_t)}  - m^2 c_t^2     \left[ (1-p_t ) c_t x   \right]^{2 p_t /(1-p_t)} .
\end{align}
Collecting the powers of $x$, we write $V$ in the simpler form as
\begin{align} \label{Vforx}
    V = \frac{\nu^2-\frac14}{x^2} - \eps\, x^{\alpha-2}
      + O \left( m^2 x^{2p_t/(1-p_t)} \right) ,
\end{align}
where
\begin{align}
    \nu^2 - \frac14 \equiv -\frac{P(2-2p_t-P)}{4(1-p_t)^2}
    \quad\Longrightarrow\quad
    \nu  = \frac{|1-p_t-P|}{2(1-p_t)} , \nn\\
    \alpha \equiv \frac{2(1-p_i)}{1-p_t} , \qquad
    \eps \equiv \frac{c_t^2}{c_1^2} \left[ (1-p_t) c_t \right]^{\alpha-2} k^2 .
\end{align}
WKB method breaks down near the singularity since $\frac{1}{x^2}$ diverges as $x \to 0$ and $V \gg \omega^2$. $p_i, p_t<1$ implies $\alpha>0$, so the second term $x^{\alpha-2}$ is less divergent than $x^{-2}$. The momentum $k$ enters only through the second term $\eps \,x ^{\alpha-2}$.

The only scale in $V$ is $x$ itself, so the $1/x^2$ term competes with $\omega^2$ at $x \sim 1/\omega$. By setting $y=\omega x$ and dividing \eqref{waveeqforx} by $\omega^2$, we obtain
\begin{align}  \label{eq:local}
    \frac{d^2\Psi}{dy^2}
    + \left[ 1 - \frac{\nu^2 - \frac14}{y^2} + \hat\eps\, y^{\alpha-2} \right] \Psi = 0 ,
    \qquad
    \hat\eps \equiv \frac{\eps}{\omega^{\alpha}} = \kappa\,\frac{k^2}{\omega^{\alpha}} ,
    \qquad
    \kappa \equiv \frac{c_t^2}{c_1^2}\left[(1-p_t)c_t\right]^{\alpha-2} .
\end{align}
The mass term in $V$ becomes $m^2\omega^{-2/(1-p_t)}y^{2p_t/(1-p_t)}$ and vanishes as $\omega \to \infty$. So we drop it.

For $y = \omega \, x \gg 1$, WKB applies to \eqref{eq:local}, and we define $\Psi_{\rm in}$ and $\Psi_{\rm out}$ as the solutions that behave as
\begin{align} \label{eq:inandout}
    \Psi_{\rm in, out} \simeq \sqrt{\frac{2}{\pi}}\, e^{\pm i \eta(y)} , \qquad y\gg1,
\end{align}
up to corrections of order $1/y^2$ and $\hat\eps\,y^{\alpha-2}$ in the amplitude, with the phase
\begin{align}  \label{eq:eta}
    \eta(y) &= \int^y\sqrt{1-\frac{\nu^2-\frac14}{y'^2}+\hat{\eps}\,y'^{\alpha-2}}\;dy' \nn\\
    &=y-\frac{(2\nu+1)\pi}{4}+\begin{cases}\dfrac{\hat{\eps}\,y^{\alpha-1}}{2(\alpha-1)} & \alpha\neq1\\[8pt] \dfrac{\hat{\eps}}{2}\log y & \alpha=1\end{cases}
    +O\left(\hat{\eps}^2,\,1/y\right).
\end{align}
The factor $\sqrt{2/\pi}$ and the constant in $\eta$ are chosen so that $\Psi_{\rm in,out}=\sqrt y\,H^{(1,2)}_\nu(y)$ at $\hat\eps=0$. Since $y=\omega x=-\omega r_*$, $\Psi_{\rm in}\simeq e^{iy}$ continues the infalling wave $e^{-i\omega r_*}$, and $\Psi_{\rm out}\simeq e^{-iy}$ continues $e^{+i\omega r_*}$, which becomes the outgoing wave at the boundary.

By the prescription of Section~\ref{sec:3}, $\Psi_{\rm sd}$ is the solution of \eqref{eq:local} that decays as $|y|\to\infty$ in the sector $-2\pi<\arg y<-\pi$. This condition, like the normalization \eqref{eq:inandout} of $\Psi_{\rm in,out}$, involves no parameter other than $\nu$, $\alpha$ and $\hat\eps$, and the normalization of $\Psi_{\rm sd}$ cancels in $R$. Since $\alpha$ is fixed by the background, we suppress it and simply write
\begin{align}
    R=R(\nu, \hat{\eps}) .
\end{align}
The important thing is that $R$ depends on the momentum $k$ only through the combination $ \hat{\eps} \propto  \frac{k^2}{\omega^\alpha}$. We do not obtain the explicit form of $R$ when $\vec{k} \neq 0$. For $k=0$, i.e., $\hat{\eps}=0$, we know $R$ because \eqref{eq:local} is Bessel's equation and the connection formula of the Hankel functions gives
\begin{align} \label{eq:R(0)}
    R(\nu, 0) = -2 \cos (\pi \nu)\, e^{- i \pi \nu },
\end{align}
with $\Psi_{\rm in,out}$ normalized to $\sqrt{y}\,H^{(1,2)}_\nu(y)$. Here we used that the solution decaying in $-2\pi<\arg y<-\pi$ is $\sqrt y\,H^{(1)}_\nu(ye^{2\pi i})$, and $H^{(1)}_\nu(ye^{2\pi i})=-H^{(1)}_\nu(y)-2\cos(\pi\nu)e^{-i\pi\nu}H^{(2)}_\nu(y)$.

We now check that, for the Schwarzschild black brane with $3\le d\le6$, the solution that vanishes at $r_e$ has this property. Near the singularity $x\propto r^{d-1}$, so $\arg y=(d-1)\arg r$, and the ray from $r=0$ towards $r_e=r_he^{-2\pi i/d}$ has $\arg y=-2\pi(d-1)/d$, which lies in $(-2\pi,-\pi)$. Along this ray, $|e^{-i\omega r_*}|$ decreases monotonically towards $r_e$ while $|e^{+i\omega r_*}|$ grows, so the solution that vanishes at $r_e$ is the one that decays along the ray.

To summarize, in the Kasner region $x\ll1/r_h$,
\begin{align} \label{Psi_sdinKasner}
    \Psi_{\rm sd} \propto \Psi_{\rm in} - R (\nu, \hat{\eps} )\, \Psi_{\rm out},
\end{align}
where $\Psi_{\rm in,out}$ are the solutions of \eqref{eq:local} fixed by their behavior \eqref{eq:inandout} at $y\gg1$. This holds throughout the Kasner region.

\subsection{Phase away from the singularity}
\label{sec:phase}

WKB applies to \eqref{eq:schr} where $\omega^2\gg|V|$, with $V=V_0+k^2f$. Near the singularity, $V\simeq(\nu^2-\frac14)/x^2$, so WKB requires $x\gg1/\omega$. Near the boundary, $V\simeq\left[(\Delta-\frac d2)^2-\frac14\right]r^2+k^2$, so it requires $r\ll\omega$ and $k\ll\omega$. We assume $k\ll\omega$ from now on. In between, $V$ is bounded, and WKB holds at large $\omega$.

In this intermediate region, the WKB solutions that continue $\Psi_{\rm in}$ and $\Psi_{\rm out}$ are
\begin{align}\label{eq:etatilde}
    \Psi_{\rm in,out}\simeq\sqrt{\frac2\pi}\left(\frac{\omega^2}{\omega^2-V}\right)^{1/4}e^{\pm i\tilde\eta(r_*)},\qquad
    \tilde\eta(r_*)=\eta(y_0)-\int_{r_*^{(0)}}^{r_*}\sqrt{\omega^2-V}\;dr_*' ,
\end{align}
where $r_*^{(0)}=-x^{(0)}$ is a point in the overlap $1/\omega\ll x\ll1/r_h$, and $y_0\equiv\omega x^{(0)}$. In the overlap, $\omega^2-V\simeq\omega^2\left[1-\frac{\nu^2-1/4}{y^2}+\hat\eps\,y^{\alpha-2}\right]$ and $dr_*=-dy/\omega$, so $\tilde\eta(r_*)=\eta(y)$ and \eqref{eq:etatilde} agrees with \eqref{eq:inandout}. In particular, $\tilde\eta$ does not depend on the choice of $x^{(0)}$.

We now continue \eqref{Psi_sdinKasner} along the real axis, passing below the horizon, to the boundary. Along this path $\omega^2-V$ does not vanish, i.e.\ there is no turning point, so each WKB solution in \eqref{eq:etatilde} continues to itself: $\Psi_{\rm in}$ and $\Psi_{\rm out}$ do not mix, and $\Psi_{\rm out}/\Psi_{\rm in}\propto e^{-2i\tilde\eta(r_*)}$. Near the boundary, WKB fails for $r\gtrsim\omega$, and there we use the Hankel functions of \eqref{eq:sdbdy} instead. Both descriptions hold in the overlap $r_h\ll r\ll\omega$. There $f\simeq h\simeq1$, $r_*\simeq r_*(\infty)-1/r$, and $V_0$ is negligible compared with $\omega^2$, so
\begin{align}
    \tilde\eta(r_*)\simeq\tilde\eta_\infty+\frac{\sqrt{\omega^2-k^2}}{r},\qquad
    \tilde\eta_\infty\equiv\eta(y_0)-\int_{r_*^{(0)}}^{r_*(\infty)}\sqrt{\omega^2-V}\;dr_* .
\end{align}
Since $H^{(1,2)}(z)\propto e^{\pm iz}$ at large $z$, this gives $\Psi_{\rm in}\propto e^{i\tilde\eta_\infty}H^{(1)}$ and $\Psi_{\rm out}\propto e^{-i\tilde\eta_\infty}H^{(2)}$, with $z=\sqrt{\omega^2-k^2}/r$. Comparing with \eqref{eq:sdbdy}, we find $e^{iS}\propto e^{-2i\tilde\eta_\infty}$, that is,
\begin{align} \label{eq:Ssplit}
    S=2\int_{r_*^{(0)}}^{r_*(\infty)}\sqrt{\omega^2-V}\;dr_*-2\eta(y_0)+{\rm const}.
\end{align}
Here, in $\tilde\eta_\infty$ and \eqref{eq:Ssplit}, $V_0$ is dropped from the integrand for $r\gg r_h$. Otherwise $V_0\simeq\big[(\Delta-\frac d2)^2-\frac14\big]r^2$ would make the integral diverge at the boundary, and its effect there is already contained in the order $\Delta-\frac d2$ of the Hankel functions.

At $k=0$ and leading order in $1/\omega$, the integral in \eqref{eq:Ssplit} simply becomes $\omega  ( r_*(\infty)-r_*^{(0)}  )$ and $\eta(y_0) \simeq y_0=-\omega r_*^{(0)}$, so \eqref{eq:Ssplit} becomes
\begin{align}
    S=2\omega\,r_*(\infty)+{\rm const}=\omega t_c+{\rm const},
\end{align}
by \eqref{eq:tc}. Since ${\rm Im}\,t_c=\beta/2$, the reflected wave is exponentially small, $|e^{iS}|\propto e^{-\beta\omega/2}$. It remains to find the $k$-dependence of $S$.

For $k\ll\omega$,
\begin{align}
    \sqrt{\omega^2-V_0-k^2f} = \sqrt{\omega^2-V_0} - \frac{k^2}{2\omega}\,f + O\left(\frac{k^2fV_0}{\omega^3},\,\frac{k^4f^2}{\omega^3}\right),
\end{align}
so that
\begin{align}  \label{eq:S_WKB}
    2 \int_{r_*^{(0)}}^{r_*(\infty)} \sqrt{\omega^2 -V }\; dr_*
    = \text{($k$-independent)} -\frac{k^2}{\omega} \int_{r_*^{(0)}}^{r_*(\infty)} f\, dr_* + \cdots .
\end{align}
The integral of $f$ converges at the boundary, since $f\to1$ and $r_*(\infty)$ is finite. The omitted terms in \eqref{eq:S_WKB} come from the next terms in the expansion of the square root, of order $k^2fV_0/\omega^3$ and $k^4f^2/\omega^3$. Near the singularity, their integrals depend on the lower end $x^{(0)}$, but these pieces cancel against the corresponding higher-order terms of $-2\eta(y_0)$, in the same way as for the leading term below. What remains is suppressed by $k^2/\omega^3$ and $k^4/\omega^3$, which vanish as $\omega\to\infty$ for $k\lesssim\omega^{1/2}$, the largest momenta that we need in Section~\ref{sec:integral}.

The $k^2$ term of \eqref{eq:S_WKB} depends on the lower end $r_*^{(0)}=-x^{(0)}$ of the integral, which lies in the Kasner region. There, the $k^2$ term of \eqref{Vforx} is $k^2f=-\eps\,x^{\alpha-2}$, so we have $- f=\kappa\,x^{\alpha-2}$. With $dr_*=-dx,$ the dependence on $x^{(0)}$ is
\begin{align} \label{eq:partofS_WKB}
    -\frac{k^2}{\omega}\int_{r_*^{(0)}}f\,dr_*\supset\frac{k^2\kappa}{\omega}\int^{x^{(0)}}x^{\alpha-2}\,dx
    =\begin{cases}\dfrac{\hat\eps\,y_0^{\alpha-1}}{\alpha-1} & \alpha\neq1,\\[8pt] \hat{\eps} \log y_0- \hat{\eps} \log \omega & \alpha=1,\end{cases}
\end{align}
where we used $\hat\eps\,y_0^{\alpha-1}=\frac{k^2\kappa}{\omega}\left(x^{(0)}\right)^{\alpha-1}$. The $y_0$-dependent part is canceled exactly by the $\hat\eps$ term of $-2\eta(y_0)$ in \eqref{eq:eta}, as it must be, since $x^{(0)}$ is arbitrary within the overlap. For $\alpha=1$, the piece $-\hat\eps\log\omega$ remains after cancellation.

Combining \eqref{eq:eta}, \eqref{eq:S_WKB} and \eqref{eq:partofS_WKB}, we obtain
\begin{align}\label{eq:S}
    S(\omega,k)=\begin{cases}\omega t_c+\theta_0\,\dfrac{k^2}{\omega}+{\rm const} & \alpha\neq1,\\[8pt]
    \omega t_c+\left(\theta_0-\kappa\log\omega\right)\dfrac{k^2}{\omega}+{\rm const} & \alpha=1,\end{cases}
\end{align}
where
\begin{align}\label{eq:theta0}
    \theta_0=-\int_{r_*^{(0)}}^{r_*(\infty)}f\,dr_*+\begin{cases}\dfrac{\kappa\,\left(x^{(0)}\right)^{\alpha-1}}{1-\alpha} & \alpha\neq1,\\[8pt] -\kappa\log x^{(0)} & \alpha=1,\end{cases}
\end{align}
does not depend on $x^{(0)}, \,  r_*^{(0)}$. For $\alpha>1$, the value of $\theta_0$ matters in Section~\ref{sec:integral}, where it controls the Gaussian integral over $\vec k$, so we show that it is real and non-zero. In this case, the integral of $f$ converges at the singularity, and the second term in \eqref{eq:theta0} equals the part of the integral between the singularity and $r_*^{(0)}$. Adding it extends the integral to the singularity:
\begin{align} \label{eq:theta0alpha}
    \theta_0=-\int_0^{r_*(\infty)}f\,dr_*=-\int_0^\infty\sqrt{\frac fh}\,\frac{dr}{r^2}  \, \in \, \RR   \qquad(\alpha>1) .
\end{align}
Since the integrand $\sqrt{f/h}/r^2$ is positive on the whole real $r$ axis and has no pole at the horizon, $\theta_0$ is real and negative. In particular, it is non-zero.

\subsection{Momentum integral}
\label{sec:integral}

Inserting \eqref{eq:S} into \eqref{eq:Gbounce}, with the constant phase absorbed into $\mathcal N$, we obtain for $k\ll\omega$
\begin{align} \label{eq:Gk}
    G_{\rm ret} (\omega, \vec k) \supset  \mathcal N \,  \omega^{2\Delta-d} \, e^{i\omega t_c} \, e^{i\theta k^2 / \omega } \, R \left(  \nu, \, \kappa\frac{k^2}{\omega^\alpha} \right) ,
\end{align}
where $\theta=\theta_0$ for $\alpha\neq1$, and $\theta=\theta_0-\kappa\log\omega$ for $\alpha=1$. The correlator at $\vec x=0$ is
\begin{align}\label{eq:xzero}
    G_{\rm ret}(\omega, \vec{x} =0) = \int \frac{d^{d-1} k }{(2\pi)^{d-1}}\, G_{\rm ret}(\omega, \vec{k}).
\end{align}

The momentum enters \eqref{eq:Gk} through two factors. $R$ depends on $k$ through $\hat\eps=\kappa k^2/\omega^\alpha$ and changes at $k\sim\omega^{\alpha/2}$. The phase $\theta k^2/\omega$ changes finitely at $k\sim\omega^{1/2}$. When we do the momentum integral, we choose the scale $\vec{k} =W \vec{\xi} $ as
\begin{align}
    W=\omega^{\bar\alpha/2},\qquad \bar\alpha\equiv\min(\alpha,1).
\end{align}

\begin{itemize}

\item $\alpha<1$: $W=\omega^{\alpha/2}$. Then $\hat\eps=\kappa\xi^2$ and $\theta_0k^2/\omega=\theta_0\omega^{\alpha-1}\xi^2\to0$, so that
\begin{align}
    G_{\rm ret} (\omega, \vec x=0) \supset \mathcal N \, \mathcal J \, \omega^{2\Delta-d+\frac{(d-1)\alpha}{2}}  e^{i\omega t_c}, \qquad
    \mathcal J \equiv \frac{\Omega_{d-2}}{(2\pi)^{d-1}}\int_0^{ e^{i\phi/2}  \infty}R(\nu,\kappa\xi^2)\,\xi^{d-2}\,d\xi , 
\end{align}
where $\Omega_{d-2}$ is the area of the unit sphere $S^{d-2}$. For $\hat\eps\gg1$, $R$ grows along the real axis. There the reflected wave can carry an exponential $e^{i\omega s}$ with $s\neq t_c$, which contributes to other singularities. In $d=4$, complex-time singularities away from $t_c$ were found for momenta $k \propto \omega$ \cite{Jia:2025jbi, Jia:2026ryl}. The term proportional to $e^{i\omega t_c}$ is the contribution of the end point $\xi=0$. We obtain it by integrating along a ray $\arg\xi=\phi/2$ from $\xi=0$, with $\phi$ chosen so that $R$ decays along $\arg\hat\eps=\phi$.

\item $\alpha>1$: $W=\omega^{1/2}$. Then $\hat \eps = \kappa \omega^{1-\alpha} \xi^2\to0$, since $1-\alpha<0$, and therefore $R (\nu, \hat \eps ) \to R(\nu,0)$. In each direction of $\vec\xi$, we obtain $\int \frac{d\xi}{2\pi} e^{i\theta_0\xi^2} = \left( \frac{i}{4\pi\theta_0} \right)^{1/2}$. The result is
\begin{align}
    G_{\rm ret} (\omega,\vec x=0) \supset  \mathcal N \, R(\nu, 0) \left( \frac{i}{4\pi\theta_0} \right)^{\frac{d-1}{2}} \omega^{2\Delta-d+\frac{d-1}{2}} \, e^{i\omega t_c}.
\end{align}

\item $\alpha=1$: $W=\omega^{1/2}$. The integrand is $e^{i (\theta_0-\kappa \log \omega ) \xi^2 } R(\nu, \kappa\xi^2)$, whose width in $\xi$ to avoid the large oscillation is $(\kappa\log\omega)^{-1/2}\to 0$. Hence $R(\nu , \hat \eps) \to R(\nu,0)$, and the result is that of $\alpha>1$ with $\theta_0$ replaced by $\theta_0 - \kappa \log \omega$:
\begin{align}
    G_{\rm ret} (\omega, \vec x=0) \supset \mathcal N \, R(\nu, 0) \left( \frac{-i}{4 \pi \kappa \log \omega}\right)^{\frac{d-1}{2}} \omega^{ 2 \Delta-d+\frac{d-1}{2}} \, e^{i\omega t_c} . 
\end{align}

\end{itemize}

\medskip
\noindent

In all cases, we obtain
\begin{align} \label{eq:G_ret(w)}
    G_{\rm ret}(\omega, \vec{x}=0) \supset C\,\omega^{ 2 \Delta  -d +  (d-1) \bar{\alpha}/2 } e^{i\omega t_c},
\end{align}
with an extra factor $(\log \omega)^{-(d-1)/2}$ for $\alpha=1$. We assume $C\neq0$. For $\alpha\ge1$, this holds unless $\nu$ is a half-integer, by \eqref{eq:R(0)}. For $\alpha<1$, it requires $\mathcal J\neq0$.

For $\alpha \neq 1$, $G_{\rm ret}(t)$ has a singular behavior as
\begin{align}
    G_{\rm ret} (t, \vec{x}=0) \propto (t_c-t)^{ -1-  2\Delta +d - (d-1 ) \bar{\alpha}/2 } ,
\end{align}
since $\int_0^{\infty} d\omega \,  \omega^q e^{-i\omega (t-t_c)} = \Gamma(q+1) [ i(t-t_c)]^{-q-1}$, as we explained in Section~\ref{sec:2}. This gives our result \eqref{b_t:general}:
\begin{align} \label{b_t:general2}
    b_t &= (d-1) -(d-1) \bar{\alpha}/2 \nn \\
    &= (d-1) \max \left( \frac{p_i-p_t}{1-p_t} , \frac12   \right)  .
\end{align}

For $\alpha=1$, the factor $(\log\omega)^{-(d-1)/2}$ becomes a logarithm in time and $G_{\rm ret} (t)$ becomes
\begin{align}  \label{eq:alpha=1}
    G_{\rm ret} (t,\vec x=0)\propto(t_c-t)^{-(2\Delta-b_t)}\left[\log\frac{1}{t_c-t}\right]^{-\frac{d-1}{2}},\qquad b_t=\frac{d-1}{2},
\end{align}
but the power is still given by \eqref{b_t:general2}.

For the AdS-Schwarzschild black brane, one has $p_t=-\frac{d-2}{d}$ and $p_i=\frac2d$. So, our formula gives $b_t=\frac d2$ which agrees with the analytic result of \cite{Jia:2026pmv} and the numerical results of \cite{Ceplak:2024bja} for $d=4,6,8$.

We expect the same method to apply to black branes whose spatial directions have different Kasner exponents $p_j$, $j=1,\dots,d-1$. This leads us to conjecture the more general formula,
\begin{align}\label{b_t:aniso}
    b_t=\sum_{j=1}^{d-1}\max\left(\frac{p_j-p_t}{1-p_t},\frac12\right),
\end{align}
which reduces to \eqref{b_t:general} when all $p_j$ are equal. We do not have numerical values of $b_t$ for such black branes to compare it with.

\section{Checks against the numerical data}
\label{sec:5}

\cite{Ceplak:2026ard} computed $b_t$ numerically, in even $d$, for the black branes
\begin{align} \label{eq:metric}
    & ds^2 = -r^2 f(r) dt^2 + \frac{dr^2}{r^2h(r)} + r^2 d \vec{x}_{d-1}^2 ,\\
    & f(r) = 1- \sum_{j=1}^K \frac{\mu_j}{r^{2j}} , \qquad  h(r) = 1- \sum_{j=1}^L \frac{\nu_j}{r^{2j}} ,\nn
\end{align}
where $f$ and $h$ vanish at a common horizon $r=r_h$, and $\mu_K, \nu_L>0$ so that the singularity at $r=0$ is spacelike. Near the singularity, $f \sim -\mu_K/r^{2K}$ and $h \sim -\nu_L/r^{2L}$, and the Kasner exponents are given by
\begin{align}
    p_t = \frac{1-K}{L} , \qquad   p_i = \frac{1}{L}.
\end{align}
One can easily check that these Kasner exponents satisfy the vacuum condition of pure gravity $\sum_a p_a = \sum_{a}p_a^2 =1$, only when $L=K=d/2$.

We consider only $K,L\ge2$, for the following reasons. When $L=1$, we obtain $p_i=1$, which violates the condition $p_i<1$. When $K=1$, $g_{tt}=-r^2f = \mu_1- r^2 $ tends to a constant at the singularity. Therefore, radial spacelike geodesics with large energy have no turning point near $r=0$ and there is no bouncing geodesic, as for the BTZ black hole and the black brane solution in the self-dual linear axion model \cite{Grozdanov:2026cut}. In agreement with this, \cite{Ceplak:2026ard} found that the coefficients $\Lambda_n$ do not follow the large-$n$ behavior used in their fits, whenever $K=1$ or $L=1$.

For $L=K$, \cite{Ceplak:2026ard} proposed the formula of $b_t(d,K,L)$ from the numerical results, as
\begin{align} \label{b_t:exp}
    b_t( d,K, K) = \frac{d-1}{1-p_t}  = \frac{K(d-1)}{2K-1} ,
\end{align}
and conjectured that $b_t=\frac{d-1}{1-p_t}$ holds for a broad class of black branes with $f=h$. Our formula \eqref{b_t:general} reduces to their relation for such black branes, since one has $p_i-p_t=1$ for $f=h$ and $\frac{p_i-p_t}{1-p_t}=\frac{1}{1-p_t}>\frac12$.

For $L\neq K$, they fitted $b_t$ numerically for $f=1-r^{-2K}$ and $h=1-r^{-2L}$, but did not find an analytic formula. They also noted that the fits are harder in this case. Table~\ref{tab:cv3} compares their values with \eqref{b_t:general}.

For $L\ge K+2$, where $\alpha>1$, \eqref{b_t:general} gives $b_t=\frac{d-1}{2}$ independently of $K$ and $L$. All 24 numerical values agree with it within $10\%$, and 21 of them within $3\%$.

The cases $L=K+1$, where $\alpha=1$, are shown in red in Table~\ref{tab:cv3}. Their numerical values are larger than the analytic values by $2$--$17\%$. We could expect this from \eqref{eq:alpha=1}. At $\alpha=1$, the singularity is not a pure power but carries a logarithm, and correspondingly
\begin{align}
    \Lambda_n \sim a^n n^{2\Delta -b_t -1} (\log n)^{-\frac{d-1}{2}} ,
\end{align}
as $n \gg 1$. A fit of such coefficients to a pure power $n^{2\Delta-b-1}$ gives
\begin{align}
    b_{\rm fit}\simeq b_t+\frac{d-1}{2\log n}.
\end{align}
This is always larger than $b_t$, growing with $d$ and decreasing with $n$. For $n \approx 400$, this shift is about $\frac{d-1}{12}$ which has the same sign and order as the observed deviations.

For $L<K$, where $\alpha<1$ and \eqref{b_t:general} depends on $p_i$, 26 of the 40 numerical values agree within $10\%$. The agreement is best in $d=4$ and $d=6$, where 9 and 8 among 10 values for each agree within $10\%$. The values that deviate by more than $10\%$ are shown in blue in Table~\ref{tab:cv3}; half of them are in $d=2$. These deviations may be due to the oscillations of $\Lambda_n$ noted in \cite{Ceplak:2026ard}, which make the fits converge slowly.

\begin{table}[H]
\centering
\small
\setlength{\tabcolsep}{4pt}
\begin{tabular}{|c|c||c|c||c|c||c|c||c|c|}
\hline
\multicolumn{2}{|c||}{} & \multicolumn{2}{c||}{$d=2$} & \multicolumn{2}{c||}{$d=4$} & \multicolumn{2}{c||}{$d=6$} & \multicolumn{2}{c|}{$d=8$} \\
\cline{3-10}
$K$ & $L$ & \begin{tabular}[c]{@{}c@{}}$b_t$\\(numeric)\end{tabular} & \begin{tabular}[c]{@{}c@{}}$b_t$\\(analytic)\end{tabular} & \begin{tabular}[c]{@{}c@{}}$b_t$\\(numeric)\end{tabular} & \begin{tabular}[c]{@{}c@{}}$b_t$\\(analytic)\end{tabular} & \begin{tabular}[c]{@{}c@{}}$b_t$\\(numeric)\end{tabular} & \begin{tabular}[c]{@{}c@{}}$b_t$\\(analytic)\end{tabular} & \begin{tabular}[c]{@{}c@{}}$b_t$\\(numeric)\end{tabular} & \begin{tabular}[c]{@{}c@{}}$b_t$\\(analytic)\end{tabular} \\
\hline\hline
\textcolor{red}{2} & \textcolor{red}{3} & \textcolor{red}{0.55} & \textcolor{red}{0.500} & \textcolor{red}{1.60} & \textcolor{red}{1.500} & \textcolor{red}{2.73} & \textcolor{red}{2.500} & \textcolor{red}{4.08} & \textcolor{red}{3.500} \\
\hline
2 & 4 & 0.53 & 0.500 & 1.48 & 1.500 & 2.57 & 2.500 & 3.57 & 3.500 \\
\hline
2 & 5 & 0.51 & 0.500 & 1.51 & 1.500 & 2.5 & 2.500 & 3.53 & 3.500 \\
\hline
2 & 6 & 0.499 & 0.500 & 1.4998 & 1.500 & 2.502 & 2.500 & 3.502 & 3.500 \\
\hline
3 & 2 & \textcolor{blue}{0.86} & \textcolor{blue}{0.750} & 2.26 & 2.250 & 3.96 & 3.750 & 5.54 & 5.250 \\
\hline
\textcolor{red}{3} & \textcolor{red}{4} & \textcolor{red}{0.54} & \textcolor{red}{0.500} & \textcolor{red}{1.60} & \textcolor{red}{1.500} & \textcolor{red}{2.69} & \textcolor{red}{2.500} & \textcolor{red}{3.73} & \textcolor{red}{3.500} \\
\hline
3 & 5 & 0.51 & 0.500 & 1.53 & 1.500 & 2.566 & 2.500 & 3.593 & 3.500 \\
\hline
3 & 6 & 0.5 & 0.500 & 1.37 & 1.500 & 2.57 & 2.500 & 3.6 & 3.500 \\
\hline
4 & 2 & \textcolor{blue}{0.7} & \textcolor{blue}{0.800} & 2.4 & 2.400 & \textcolor{blue}{4.9} & \textcolor{blue}{4.000} & \textcolor{blue}{6.3} & \textcolor{blue}{5.600} \\
\hline
4 & 3 & \textcolor{blue}{1.02} & \textcolor{blue}{0.667} & 2.13 & 2.000 & 3.19 & 3.333 & 4.59 & 4.667 \\
\hline
\textcolor{red}{4} & \textcolor{red}{5} & \textcolor{red}{0.51} & \textcolor{red}{0.500} & \textcolor{red}{1.60} & \textcolor{red}{1.500} & \textcolor{red}{2.63} & \textcolor{red}{2.500} & \textcolor{red}{3.71} & \textcolor{red}{3.500} \\
\hline
4 & 6 & 0.50 & 0.500 & 1.47 & 1.500 & 2.57 & 2.500 & 3.61 & 3.500 \\
\hline
5 & 2 & 0.78 & 0.833 & 2.48 & 2.500 & 4.4 & 4.167 & \textcolor{blue}{7} & \textcolor{blue}{5.833} \\
\hline
5 & 3 & \textcolor{blue}{0.60} & \textcolor{blue}{0.714} & 2.0 & 2.143 & 3.4 & 3.571 & 5.2 & 5.000 \\
\hline
5 & 4 & \textcolor{blue}{0.85} & \textcolor{blue}{0.625} & 2.04 & 1.875 & 3.26 & 3.125 & 4.15 & 4.375 \\
\hline
\textcolor{red}{5} & \textcolor{red}{6} & \textcolor{red}{0.55} & \textcolor{red}{0.500} & \textcolor{red}{1.6} & \textcolor{red}{1.500} & \textcolor{red}{2.62} & \textcolor{red}{2.500} & \textcolor{red}{3.60} & \textcolor{red}{3.500} \\
\hline
6 & 2 & 0.862 & 0.857 & 2.583 & 2.571 & 4.4 & 4.286 & \textcolor{blue}{8} & \textcolor{blue}{6.000} \\
\hline
6 & 3 & \textcolor{blue}{0.52} & \textcolor{blue}{0.750} & 2.39 & 2.250 & \textcolor{blue}{4.54} & \textcolor{blue}{3.750} & \textcolor{blue}{6} & \textcolor{blue}{5.250} \\
\hline
6 & 4 & 0.63 & 0.667 & \textcolor{blue}{1.79} & \textcolor{blue}{2.000} & 3.04 & 3.333 & 4.35 & 4.667 \\
\hline
6 & 5 & \textcolor{blue}{0.78} & \textcolor{blue}{0.600} & 1.90 & 1.800 & 3.04 & 3.000 & 4.24 & 4.200 \\
\hline
\end{tabular}
\caption{Numerical values of $b_t$ for $\Delta=3/2$ and $K\neq L$, from Table~III in \protect{\cite{Ceplak:2026ard}}. Red: $L=K+1$, i.e.\ $\alpha=1$, where the singularity carries a logarithm \eqref{eq:alpha=1}. Blue: other cases that deviate from the analytic formula by more than $10\%$.}
\label{tab:cv3}
\end{table}

Our formula \eqref{b_t:general} fits the data much better than the extension $b_t=\frac{d-1}{1-p_t}$ of \eqref{b_t:exp}. For example, the latter gives $b_t=6$ for $(K,L)=(2,6)$ in $d=8$, and $b_t=0.857$ for $(6,2)$ in $d=4$, while the numerical values are $3.502$ and $2.583$. But our formula gives $3.5$ and $2.571$, respectively.

\section{Predictions}
\label{sec:6}

In this section, we apply \eqref{b_t:general} to black branes of gravity coupled to matter, for which no value of $b_t$ is available. We consider two families in AdS$_4$ ($d=3$): the Einstein--scalar black branes studied in \cite{Hartnoll:2026vhu} in Section~\ref{sec:EinS}, and the Einstein--Maxwell--dilaton black branes studied in \cite{Shi:2026cfq} in Section~\ref{sec:EMD}.

\subsection{Einstein--scalar black branes}
\label{sec:EinS}

\cite{Hartnoll:2026vhu} studied four different backgrounds. They are Schwarzschild-AdS$_4$, a numerical solution of an Einstein--scalar model (EinS), and two exact solutions of an Einstein--scalar model with a potential (EinSV1, EinSV2). From the Kasner exponents of each, they predicted the large-$n$ behavior of $\partial\omega_n/\partial k^2|_{k=0}$ and confirmed it numerically.

No value of $b_t$ has been computed numerically for these backgrounds. Inserting their $p_t,p_i$ into \eqref{b_t:general} gives the predictions below.

\begin{center}
\small
\renewcommand{\arraystretch}{1.25}
\begin{tabular}{lccccc}
\toprule
model (Table 1 of \protect{\cite{Hartnoll:2026vhu}}) & $p_t$ & $p_i$ & $\alpha$ & $b_t$ (analytic)    \\
\midrule
Schw & $-\frac13$ & $\frac23$ & $\frac12$ & $1.5$  \\
EinS & $-0.3172$ & $0.6586$ & $0.5184$ & $1.482$  \\
EinSV1 & $-\frac17$ & $\frac47$ & $\frac34$ & $1.25$   \\
EinSV2 & $\frac15$ & $\frac25$ & $\frac32$ & $1^\dagger $   \\
\bottomrule
\end{tabular}

\smallskip
{\footnotesize $^\dagger$ Since $p_t>0$, the existence of a bouncing singularity is not guaranteed. The value applies if it exists.}

\end{center}

The first row is the Schwarzschild value $b_t=\frac d2=\frac32$, derived for general $d$ in \cite{Jia:2026pmv}, even though it has not been checked numerically for $d=3$. For the other three rows, one can see $p_t+2p_i=1$ still holds, but $p_t^2+2p_i^2 \neq 1$, because we are considering the scalar field coupled to gravity. The scalar field also contributes to the second Kasner condition. The $b_t$ values for these three backgrounds are our predictions.

The last row has $p_t>0$. For $g_{tt}g_{rr}=-1$, a sufficient (but not necessary) condition for a bouncing geodesic is that $|g_{tt}|$ grows towards the singularity \cite{Grozdanov:2026cut}, i.e.\ $p_t<0$. So the value of $b_t$ in the last row applies only if this background has a bouncing geodesic.

\subsection{Einstein--Maxwell--dilaton black branes}
\label{sec:EMD}

We consider the analytic Einstein--Maxwell--dilaton black branes in AdS$_{d+1}$ of \cite{Hendi:2010gq, Gouteraux:2011ce}, in which the dilaton $\phi$ couples to the Maxwell field through $e^{-(d-2)\delta\phi}F^2$. The coupling $\delta>0$ labels the theory. Their interiors were analyzed in \cite{Arean:2024pzo}, and their high-overtone quasinormal modes in \cite{Shi:2026cfq}. We follow the parametrization of these papers. We set $d=3$ and write $a \equiv \delta^2$. For $a<\frac13$, the black brane has an inner horizon and the singularity is timelike. For $a\ge\frac13$, the singularity is spacelike with Kasner exponents \cite{Arean:2024pzo}
\begin{align}
    (p_t, \, p_i)= \left( \frac{1-a}{1+3a}, \, \frac{2a}{1+3a} \right),
\end{align}
which satisfy $p_t+2p_i=1$, but $p_t^2+2p_i^2 \neq 1$ for any finite $a$, for the same reason as in the previous subsection. For these backgrounds, $\alpha=\frac{1+a}{2a}$ is the exponent $q$ of \cite{Shi:2026cfq} governing the momentum dependence of the high-overtone quasinormal modes.

For $a>1$, we have $p_t<0$, so $|g_{tt}|$ grows towards the singularity. These black branes have $g_{tt}g_{rr}\neq-1$, so the criterion of \cite{Grozdanov:2026cut} does not apply directly, but we expect a bouncing geodesic. Since we have $\alpha<1$ for $a>1$, our formula \eqref{b_t:general} gives
\begin{align}
    b_t=\frac{3a-1}{2a},
\end{align}
which increases from $1$ at $a=1$ to the Schwarzschild-AdS$_4$ value $\frac32$ as $a \to \infty$. In this limit, $p_t^2+2p_i^2\to1$.

For $\frac13\le a\le1$, we have $p_t\ge0$ and the existence of a bouncing singularity is not established. If it exists, our formula gives $b_t=1$.

\section{Conclusion and discussions}
\label{sec:7}

In this paper, we have shown that the additional power $b_t$ in the bouncing singularity of the thermal correlator can be written in terms of the Kasner exponents $(p_t,p_i)$ for planar black branes in AdS$_{d+1}$, as in \eqref{b_t:general}. We obtained it from the exponentially small term of the retarded correlator at large $\omega$, by using the wave reflected at the singularity and integrating over the spatial momentum $\vec k$.

When $\alpha\equiv\frac{2(1-p_i)}{1-p_t}>1$, the phase oscillation $e^{i\theta_0k^2/\omega}$ restricts the range of the $\vec k$ integral to $k\lesssim\omega^{1/2}$. In this range the reflection coefficient is just $R(\nu,0)$, so $b_t=\frac{d-1}{2}$ is determined only by the dimension. When $\alpha<1$, the reflection coefficient $R(\nu,\hat\eps)$ already changes at the smaller scale $k\sim\omega^{\alpha/2}$, where the phase oscillation is still negligible. We integrate in this scale, so the reflection coefficient contributes, and $b_t$ contains the information of the Kasner exponents. As a special case, if $\alpha=1$, the integral of $f$ in the phase diverges logarithmically at the singularity, and the correlator carries the logarithmic factor as well as the pure power, as in \eqref{eq:alpha=1}.

We made some assumptions to derive $b_t$. First, we adopted the proposal of \cite{Afkhami-Jeddi:2025wra} that the Green's function computed from the steepest-descent solution $\Psi_{\rm sd}$ is the canonical Borel sum of the large-$\omega$ expansion. Second, we selected $\Psi_{\rm sd}$ near the singularity as the solution that decays in the sector $-2\pi<\arg y<-\pi$. For the Schwarzschild black brane with $3\le d\le6$, where the steepest-descent curve has an end point, this follows from the condition that $\Psi_{\rm sd}$ vanishes there. For $f\neq h$, where the curve has no end point, we took it as a prescription, for all $\omega$ and $k$. Third, for $\alpha<1$ we assumed $\mathcal J\neq0$. Fourth, when not all $s_n$ are even, as in odd $d$, we assumed that $G_T(-it+0^+)$ neither strengthens nor cancels the singularity at $t_c$. This is not a rigorous proof, and it would be interesting to derive our formula without these assumptions, or to construct a more general formula for $b_t$.

The numerical results of \cite{Ceplak:2026ard} support our derivation. All 24 values with $\alpha>1$ agree with \eqref{b_t:general} within $10\%$. For $\alpha=1$, all numerical values are larger than our formula, as expected from the logarithmic factor. There were some exceptions with $\alpha<1$, where the numerical value deviates from our formula by more than $10\%$. We leave open whether this is due to the lack of numerical accuracy or the failure of our formula.

We also provided the predictions of $b_t$ for the Einstein--scalar and Einstein--Maxwell--dilaton black branes, whose high-overtone quasinormal modes were studied in \cite{Hartnoll:2026vhu, Shi:2026cfq}. It would be interesting to extend our analysis to black branes whose singularity is timelike and lies behind an inner horizon, such as the Reissner--Nordstr\"om--AdS black brane, where the bouncing singularity has been studied in \cite{Ceplak:2025dds, Chakravarty:2026fdm}.

Our analysis can also be extended to interiors with different Kasner exponents in different spatial directions \cite{Gao:2023zbd}. For such interiors, we conjectured in Section~\ref{sec:integral} that
\begin{align}
    b_t=\sum_{j=1}^{d-1}\max\left(\frac{p_j-p_t}{1-p_t},\frac12\right). \nn
\end{align}
It would be interesting to compute $b_t$ numerically for these black branes.

\vskip 0.5cm

\hspace*{-0.8cm} {\bf\large Acknowledgements}
\vskip 0.2cm

\hspace*{-0.75cm} 
I thank Seok Kim for helpful discussions. This work is supported in part by the NRF grant RS-2026-25477643. I used Claude Code to produce figures and tables.

\appendix

\bibliographystyle{JHEP}
\bibliography{refs}

\end{document}